\documentclass[11pt]{article}
\usepackage{graphicx}              % for \includegraphics

\begin{document}

\begin{titlepage}
\date{}
\title{{\Large\bf Black Hole Evaporation Entails an Objective Passage of Time }}

\author{\Large Avshalom C. Elitzur$^{a}$, Shahar Dolev$^{b}$}
\maketitle
\begin{center}
$^{a}$ Department of Chemical Physics, Weizmann Institute of Science, 76100
Rehovot, Israel.\\

\noindent
E-mail: cfeli@weizmann.weizmann.ac.il

\vspace{0.1cm}

$^{b}$ The Kohn Institute for the History and  Philosophy of Sciences,  Tel-Aviv
University, 69978 Tel-Aviv, Israel.\\

\noindent
E-mail: shahar@email.com

\end{center}

\thispagestyle{empty}
\abstract\noindent

Time's apparent passage has long been debated by philosophers, with no decisive argument for or against its objective existence. In this paper we show that introducing the issue of determinism gives the debate a new, empirical twist. We prove that any theory that states that the basic laws of physics are time-symmetric must be strictly deterministic. It is only determinism that enables time reversal, whether theoretical or experimental, of anyentropy-increasing process. A contradiction therefore arises between Hawking's \cite{Haw1} argument that physical law is time-symmetric and his controversial claim \cite{Haw2} that black-hole evaporation introduces a fundamental unpredictability into the physical world. The latter claim forcibly entails an intrinsic time-arrow independent of boundary conditions. A simulation of a simple system under time reversal shows how an intrinsic time arrow re-emerges, destroying the time reversal, when even the slightest failure of determinism occurs. This proof is then extended to the classical behavior of black holes. We conclude with pointing out the affinity between time's arrow and its apparent passage.

\vspace{1cm}

{\bf Key words:} Time's Passage, Time's arrow; Determinism; Black holes, Information loss

\end{titlepage}

%%%%%%%%%%%%%%%%%%%%%%%%%%%%%%

Ever since Minkowski has formulated the geometrical basis of relativity theory, in which time is the fourth dimension, time's apparent ``passage" has presented a difficulty. For we experience events in fundamentally different ways in time than in space. In space, we can observe sequences of events in whatever order we chose (e.g., from left to right or from right to left). In time, in contrast, we experience such sequences only from past to future. In space, we can remain at the same place. In time, in contrast, moments seem to follow one another. Time thus seems to be ``moving" or ``flowing." 

A few authors indeed argue that time has an objective property of ``passage" (\cite{Hor}, \cite{Pri}, \cite{Dav}). These speculations, however, run into logical loops (higher and higher order times) and/or invoke ``absolute time" that conflicts with relativity theory \cite{God}. It is therefore understandable that most physicists prefer to regard time's passage as illusion. In the Einstein-Minkowski spacetime, all events - past, present and future - have the same degree of existence. 

The trouble with this debate is that it seems to belong entirely to the realm of philosophy. All physical observations are equally consistent with a Block Universe (``all events coexist along time") and Becoming (``events are created anew one after another, along time"). The Block Universe interpretation is awkward, implying that all future events, including our apparently-free choices, ``already" exist. Becoming, on the other hand, gives rise to the difficulties pointed out above. In the absence of a decisive empirical test, most physicists relegate the entire issue to the philosophers. Unfortunately, as the recent books by Oklander \& Smith (e.g., \cite{Oak1}, \cite{Oak2}) show, philosophy has so far very little success in dealing with this stalemate.

In this paper we show that this problem can turn into an empirical issue once the question of determinism is raised. The gist of our argument \cite{Eli} is this: {\it i)}Indeterminism in itself is as time-symmetric as determinism. {\it ii)}However, in a world in which entropy is increasing, indeterminism indicates that causality itself is asymmetric: Past events can be the causes of future events but not {\it vice versa}. {\it iii)}Hence, if future events have no causal efficacy on past events, the Block Universe interpretation is superfluous.

This proof is at the logical level. The ensuing physical question is, of course, whether our world {\it is}, indeed, indeterministic. This question will be discussed next. 

\section {Are Causal Relations Time-Symmetric?}

Very likely, time's apparent passage is closely related to its manifest asymmetry. The basic laws of physics are invariant under time reversal, yet the macroscopic world is markedly time-asymmetric. Typical time-arrows known, {\it inter alia}, from thermodynamics, electromagnetism, cosmology, particle physics and black-hole physics. How they are related to one another is the subject of a continuing controversy (e.g., \cite{Dav2}, \cite{Hal}, \cite{Sav}, \cite{Zeh}). 

Hawking \cite{Haw2} and Penrose \cite{Pen} have long been taking opposing views on this issue, their debate being recently published as a book \cite{Haw3}. Penrose believes that when physics attains the long-desired theory of quantum gravity, that theory will reveal an intrinsic time-asymmetry in the basic laws of physics. Hawking, in contrast, argues that the observed arrows of time only reflect some unique initial conditions in the universe's evolution (\cite{Haw3}, p. 8). Causation itself, he stresses, is perfectly time-symmetric: 

So if state A evolved into state B, one could say that A caused B. But one could equally well look at it in the other direction of time, and say that B caused A. So causality does not define a direction of time (\cite{Haw1}, p. 346).  

On one point, however, both adversaries agree. For reasons revealed long ago by Hawking (\cite{Haw4}, \cite{Haw5}), black holes must eventually evaporate in the form of purely thermal radiation. Now, both Penrose and Hawking agree that all the information about the objects that have fallen into the black hole is destroyed when the black hole evaporates. Unlike the ordinary loss of information due to mixing or noise (which can, in principle, be retrieved), this information loss by black hole evaporation is absolute. ``Information annihilation" would therefore be a better term. In Hawking's words: ``quantum gravity introduces a new level of unpredictability into physics over and above the uncertainty usually associated with quantum theory" (\cite{Haw3}, p. 60).

We would like to show that Hawking's two assertions are mutually incompatible. If information is indeed annihilated, by whatever process, then time's arrow is inherent to causality itself. 

It should be stressed that we do not attempt to endorse Hawking's information-annihilation hypothesis but only point out its surprising consequences. Our conclusion, however, rigorously applies to any other theory that makes a similar assumption, be it the GRW model of spontaneous collapse \cite{Ghi}, Penrose's \cite{Pen} hypothesis of the role of gravity in quantum measurement, or any other assertion that information is annihilated at the quantum level.

\section {The Paradox: Could we be Living in a Thermodynamically-Inverted Universe?}

We begin by pointing out an observational paradox. Could our universe's entropy be decreasing rather than increasing? Much as this possibility sounds absurd, it cannot be ruled out. The reason for this was given long ago by Hawking \cite{Haw5}: {\it Have we lived in an entropy decreasing system, we would not be able to notice it}. The time orientation of all biological processes (as we have shown elsewhere in detail (\cite{Eli2}, \cite{Dol})) relies solely on entropy's increase. Consequently, our memory and our distinction between ``past" and ``future" are similarly oriented. Have the Universe's entropy been decreasing, memory and cognition would run backwards too and the notions ``past" and ``future" would accordingly be inverted. It therefore cannot be ruled out that we actually live in a universe whose entropy is decreasing.

One might dismiss this possibility by arguing that such unique initial conditions that lead to entropy decrease are highly improbable. Unfortunately, when this argument is applied to the entire universe it becomes circular. Consider the following statement:

\begin{quote}

{\it {\bf A:} It is improbable for a system to have such precise correlations in its initial state that would eventually result in later entropy decrease. }

\end{quote}

This perfectly accords with everyday experience. And since this is a probabilistic statement, it can further be rephrased quantitatively: 

\begin{quote}

{\it {\bf A1:} The larger the system, the more improbable it is to have such precise correlations in its initial state. }

\end{quote}

Which, again, perfectly accord with both empirical evidence and probability laws. What is more natural, then, than to generalize this statement to the ultimate system, the entire Universe? Our above innocuous statement would then read:

\begin{quote}

{\it {\bf A2:} It is most improbable for the Universe to have such precise correlations in its initial state that eventually result in later entropy decrease. }

\end{quote}

But this statement, surprisingly, has a fatal flaw: It employs the temporal designation ``initial state" and ``later." {\it But what are ``past" and ``future" other than the directions in which entropy is, respectively, low and high?} When one observes a system's evolution, one assigns the temporal designations ``before" and ``after" according to the universal time arrow prevailing outside the system. No such external reference clock exists, of course, for the Universe itself. Statement {\bf A2} is therefore a mere tautology!

For this reason, most physicists have adopted the ``atemporal" account of entropy increase \cite{Pri2}. Instead of ``initial" and ``final" conditions, the atemporal account employs the neutral term ``boundary conditions." One boundary of the universe's evolution has low entropy while the other boundary's entropy is high. Each state can be viewed as either the ``cause" or ``effect" of the other. Hawking, as noted above, employs this atemporal interpretation, yet Price \cite{Pri3} has criticized him for not employing it systematically enough. Wheeler \& Feynmann \cite{Whe} employed this atemporal account as early as in 1949: ``The stone hit the ground because it was dropped from a height. Equally well: the stone fell from a height because it was going to hit the ground." 

Is there any other way out of the Block Universe? We next show that the atemporal account is valid only if one assumes absolute determinism. If determinism ever fails even slightly, the Second Law would turn out to be the result of a fundamental time asymmetry underlying causality itself. Then, if causality turns out to be time-asymmetric, time's passage might be real after all.

\section {Indeterminism Entails a Universal Time-Arrow}

There is a well-known yet crucial difference between the normal, entropy increasing evolution, and the time-reversed, order increasing one. {\it The latter, not the former, requires infinitely precise pre-arrangements of all the system's elementary particles}. For a normal process, no special care is needed to arrange its particles so as to increase entropy. Boltzmann's definition of entropy, $S=k \ln W$, is based on the fact that there are numerous microscopic arrangements that make disordered states (e.g. thermal equilibrium) but only few arrangements that make ordered ones. Hence, nearly every initial arrangement will eventually give rise to entropy increase. Consequently, a change in the initial arrangements can hardly affect the increase of entropy (Fig. 2a). 

The situation is radically different with the time-reversed system: The slightest change in the position or momentum of a single particle will create a disturbance in the system's evolution that - given sufficiently many interactions between the particles - will further increase as the system evolves. Consequently, entropy will increase in the time-reversed system too (Fig. 2b). As Yakir Aharonov so vividly puts it, take out one worm from a dead person's grave, and the time-reversed evolution will fail to bring him or her back to life. 

\begin{figure}
\centering
\includegraphics[scale=0.95]{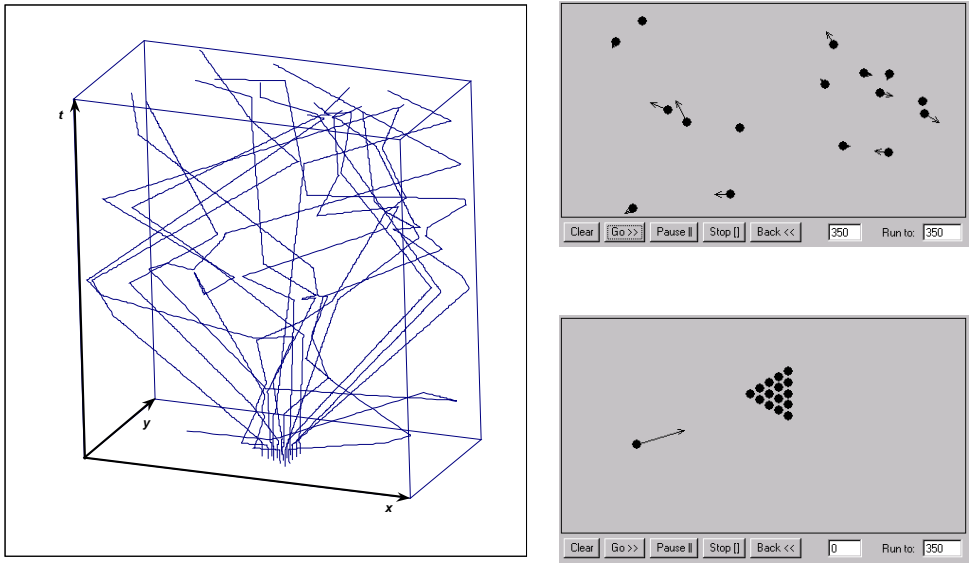}
\renewcommand{\thefigure}{1a}
\caption{A computer simulation of an entropy increasing process, with the initial and final states (right) and the entire process using a spacetime diagram (left). One billiard ball hits a group of ordered balls at rest, dispersing them all over the table. After repeated collisions between the balls, the energy and momentum of the first ball is nearly equally divided between the balls.}
\end{figure}

\begin{figure}
\begin{center}
\includegraphics[scale=0.95]{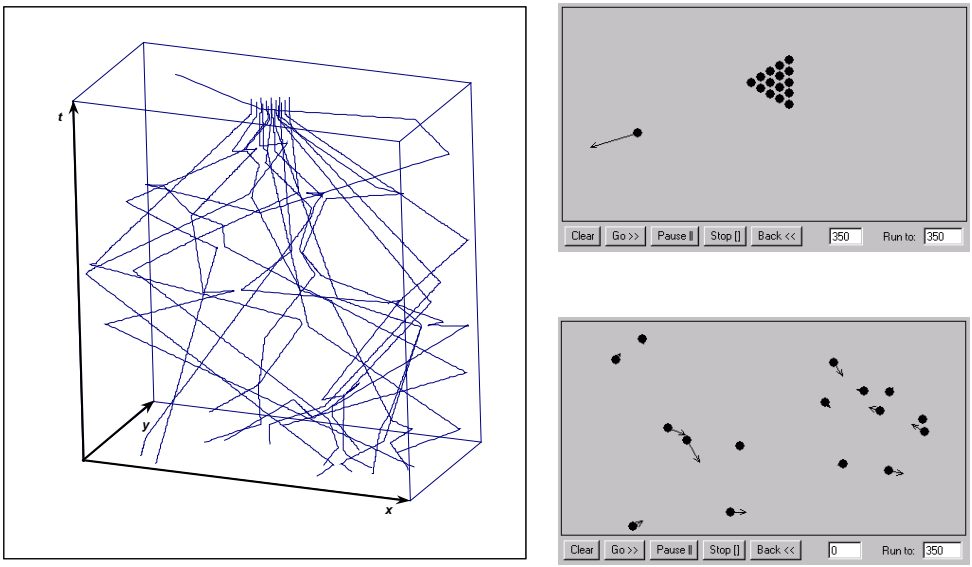}
\renewcommand{\thefigure}{1b}
\caption{The time-reversed process. All the momenta of the balls are reversed at {\it $t_{350}$}. Eventually, the initial ordered group is re-formed, as at {\it $t_{0}$}, ejecting back the first ball.}
\end{center}
\end{figure}

\begin{figure}
\begin{center}
\includegraphics{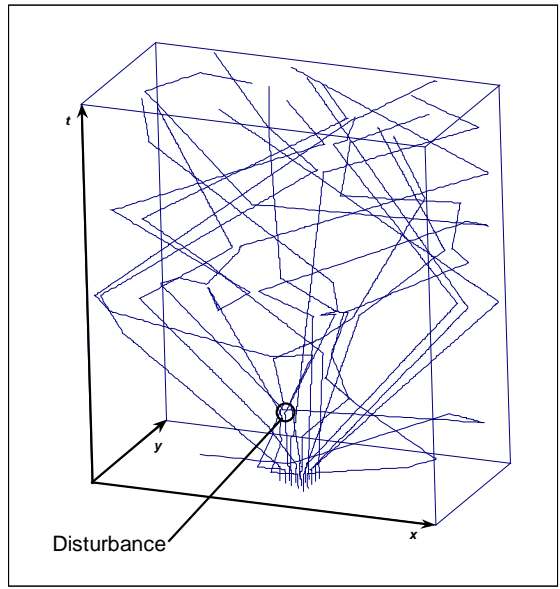}
\renewcommand{\thefigure}{2a}
\caption{The same simulation as in Fig. 1a, with a slight disturbance in the trajectory of one ball (marked by the small circle). Entropy increase seems to be indistinguishable from that of Fig 1a. }
\end{center}
\end{figure}

\begin{figure}
\begin{center}
\includegraphics{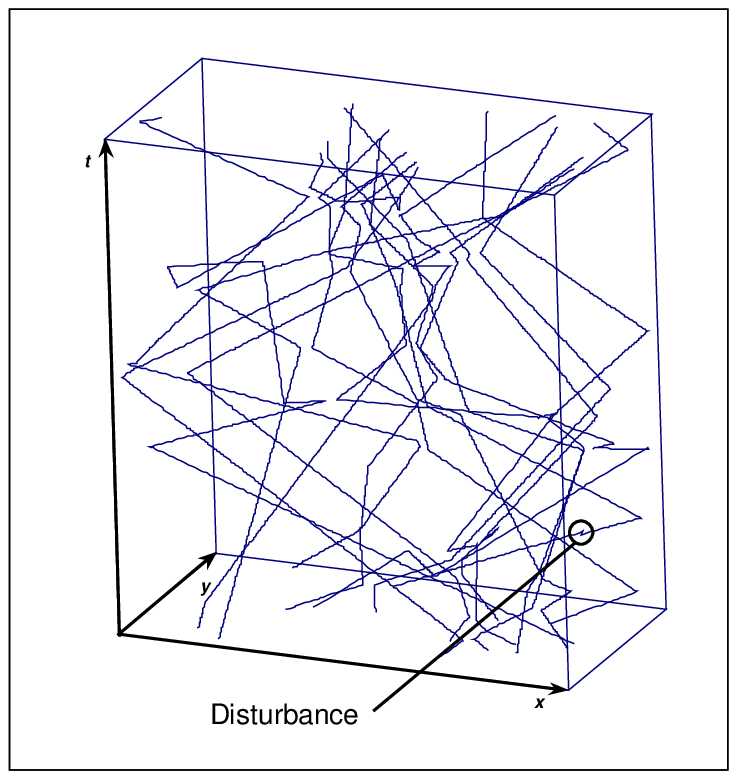}
\renewcommand{\thefigure}{2b}
\caption{The same computer simulation as in Fig. 1b, with a similar disturbance. Here, the return to the ordered initial state fails. }
\end{center}
\end{figure}

This restriction is almost trivial, but its far-reaching consequences have not been explored. Had physics been able to prove that determinism does not always hold - that some processes are governed by fundamentally probabilistic laws - it would follow that entropy {\it always} increases, regardless of the system's initial conditions. An intrinsic time-arrow would then emerge in {\it any} system, independent of the initial conditions. The emerging time arrow would be congruent with that of the entire universe, of which closed systems are supposed to be shielded. 

\section {Quantum Mechanics does Not  Disprove Determinism}

At first sight, quantum mechanics seems to have disproved determinism long ago, thereby giving an intrinsic time arrow. Indeed, several authors have shown that in order to preserve Lorentz invariance, QM must contain a genuinely stochastic element \cite{Bel}. 

Has there been a definite proof for this assertion, we would be able to rest our case at this point. At present, however, QM has not yet ruled out the possibility that determinism still exists at some unobservable level (\cite{Eli3}, \cite{Eli4}). 

The indeterminacy of QM is ascribed to the abrupt change caused by measurement. However, the alleged irreversibility of measurement has never been rigorously proved. To see this, let us examine an alleged example of indeterminism often used by Penrose (e.g., \cite{Pen}, \cite{Pen2}). A half-silvered mirror splits the wave function of a single photon such that one half hits a detector and the other half hits the wall. In 50\% of the cases, the detector will click. Suppose now that we time-reverse this process. Would the photon ejected from the detector return back to the source? Not always: The photon's wave function will be split again by the beam splitter when traversing its path in the reverse direction, giving 50\% probability that the photon will not return to the lamp but rather go to the opposite wall. This, for Penrose, indicated that there is something essentially time-asymmetric at the quantum level. 

Recently, however, Penrose \cite{Haw3} conceded that this asymmetry may merely be reflecting the asymmetry of the boundary conditions. In the normal time-evolution, half of the wave function initially goes to the wall. Now this half wave function is a real physical phenomenon: Reflected back by a mirror, this half can be used to create interference effects (The Elitzur-Vaidman bomb-testing experiment \cite{Eli5}, to which Penrose \cite{Pen2} gives a vivid exposition, proves how real this ``empty" half is). Therefore, for a real time reversal to take place, the wall's absorption of the half wave-function (i.e., its interaction-free measurement) must be reversed too. Once we took care to make the time reversal thus complete, the photon would indeed return to the lamp from which it has been initially emitted. Penrose, of course, believes that this would not happen, for he regards ``collapse of the wave function" as a real process. However, this remains a hypothesis. No experiment is known today that can favor this interpretation over other, time-symmetric ones. For example, the ``guide wave" or the ``many worlds" interpretations assume that some hidden variables, in the form of empty waves or parallel universes, remain after the measurement, preserving all the seemingly-lost information (see Unruh \cite{Unr} for another objection to Penrose's argument).  

Quantum mechanics, therefore, does not give an intrinsic time-arrow. It allows deterministic processes with (non-local) hidden variables. Hence, like classical physics, quantum theory allows any process to be time-reversed under the appropriate initial conditions.

\section {Black Holes' Evaporation Undermines Determinism}

It is black holes, however, that seem to provide what we are looking for. A black hole is unique in that it ``compresses" all the degrees of freedom of any object that it has swallowed into three, namely, mass, charge and angular momentum. Classically, the lost information about these objects could still exist somehow within the black hole, but Hawking has shown that this is not the case. Due to the Bekenstein-Hawking effect (\cite{Bek}, \cite{Haw4}, \cite{Haw5}), black holes eventually evaporate in the form of particles whose mass distribution is that of black body radiation at the black hole's temperature. Since this radiation results from quantum fluctuations of the vacuum near the black hole's event horizon, it seems to be absolutely thermal, being unrelated to, and preserving nothing of the black hole's content. This gives rise to entropy that is not due to coarse graining, as in classical physics. Rather, the entropy seems to be absolute. 

Naturally, Hawking's claim has met widespread objections, yet none turned out to be decisive. Preskill \cite{Per}, initially one of the opponents, has thoroughly reviewed all the proposals to avoid the information loss paradox (some of which being as desperate as proposing that the information goes to other universes) and found all of them deficient (see also \cite{Eli}). ``The information loss paradox," he concludes, ``may be a genuine failing of $20^{th}$ Century physics, and a signal that we must recast the foundations of our discipline."

In view of our preceding analysis, it is obvious that, if Hawking's argument is sound, then its most immediate bearing, unnoticed so far, is on the origin of time symmetry. 

Consider, then, the following thought experiment. Let a closed system undergo a normal evolution, whereby its entropy will increase with time. However, let the system have enough mass and time that will allow a black hole to form and evaporate. Opening the system after sufficient time and inspecting its entropy, we find that its entropy has increased. This is not surprising: If Hawking's hypothesis is correct, the particles into which the black hole has evaporated could not preserve the positions and momenta of the objects swallowed earlier by the black hole. Hence, the black hole has merely added to the system's entropy, in a way similar to that of the single event in Fig. 2a where the causal chain was interfered with. 

Consider next the time-reversed system. Let there be a similar closed system, with the positions and momenta of all its particles pre-correlated with utmost precision such that its entropy would decrease with time. And here too, the amount of matter and the time allocated to the system suffice for the formation and evaporation of a black hole. Opening the system at the end of the experiment, we find that time-reversal has failed: Entropy has increased in this case too. 

The reason is clear: The black hole's information annihilating effect has ruined the pre-arranged correlations with which the initial state has been prepared. This parallels the simplified case shown in Fig. 2b, with the difference that the failure of determinism entailed by the black hole effects not only one but {\it numerous} particles.\footnote{ Hawking \cite{Haw3} argues against an intrinsic time-arrow by employing a thought experiment (p. 124) that involves a closed box containing only photons at maximum entropy. Indeed, a black hole formed under such conditions would not create a time arrow (the final state would be, again, thermal photons in maximum entropy...). This example, however, involves a system that has no time arrow to begin with, therefore not applying to our case. In particular, Hawking's example fails to explain the following time-asymmetry: Given an initial arrow of time in a closed system, a black hole only strengthen this local arrow if it accords with the universal arrow, but reverses it when the local arrow points to the opposite direction.} 

Let us put this failure in physical terms. Preparing a black hole in a time-reversed system amounts to preparing a {\it white hole}, i.e., a singularity from which macroscopic objects are ejected. But this is precisely what we cannot do. In order to create a white hole, numerous particles must be directed towards one point (this is the time-reversal of the black hole's evaporation). With a sufficiently huge number of particles a singularity will indeed form, which will later evaporate. However, when it evaporates, it evaporates again into particles whose spectrum is {\it thermal}; no objects with complex physical attributes can emerge from it. Although Hawking seems to be unaware of it, his argument of information annihilation provides a perfect explanation for the absence of white holes!

A normal system, then, increases its entropy when an information-annihilating event is formed within it, but so does a time-reversed system. The conclusion therefore follows: {\it In any closed system whose evolution contains an information-annihilating event, this event gives rise to an intrinsic time-arrow that disregards the system's boundary conditions, but complies with the time arrow of the universe, of which closed systems are supposed to be shielded} (Fig. 3).

\begin{figure}
\begin{center}
\includegraphics{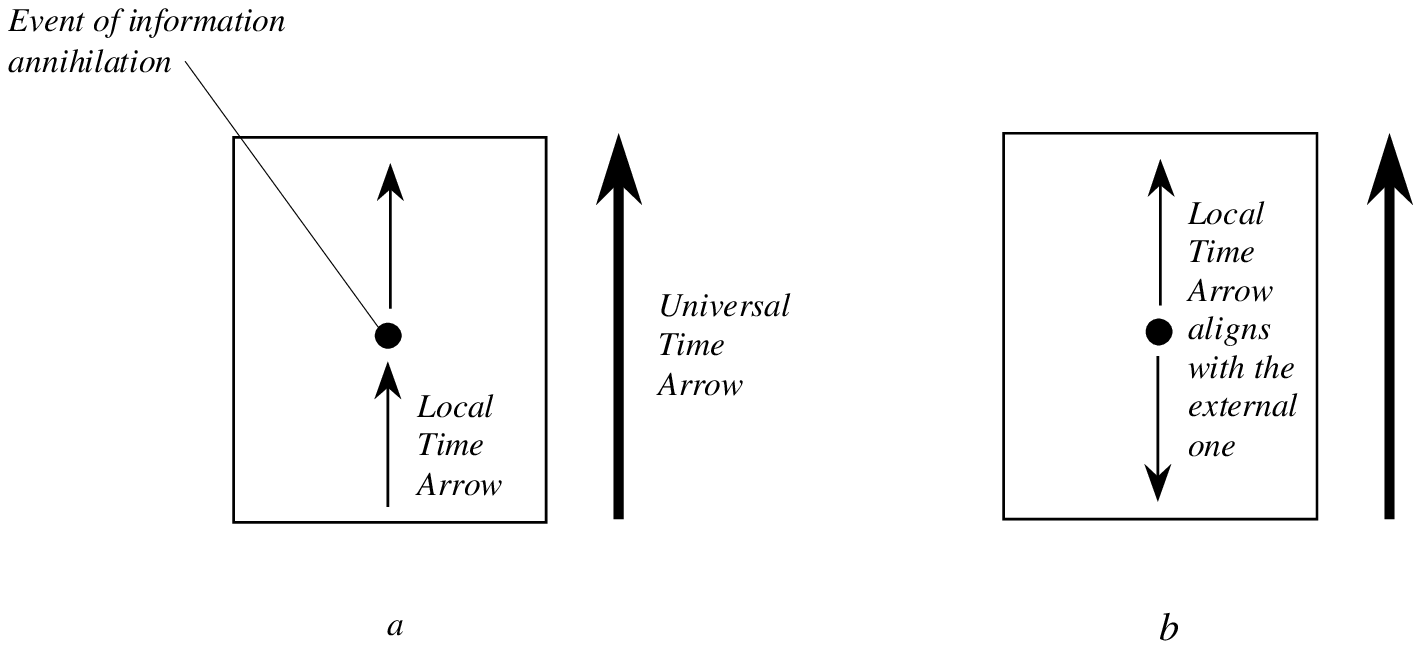}
\renewcommand{\thefigure}{3}
\caption{When an information-annihilating event occurs within a closed system, then, if the system's internal time-arrow accords with that of the universe (a), the event leaves the internal arrow unaffected. However, if the internal time-arrow is opposite to that of the universe (b), the event reverses it back.}
\end{center}
\end{figure}

\section {A Black Hole gives rise to a Time Arrow Even if it Preserves Information}

So far, we have taken Hawking's information-annihilation hypothesis as a possible source for an intrinsic time-arrow. However, a closer inspection shows that, even without assuming Hawking's hypothesis, a black hole must give rise to an inherent time-arrow in any system in which it resides. 

Suppose, then, that the evaporating black hole's radiation of is not really thermal, i.e., that information about the in-falling matter somehow survives in the particles emitted by the evaporating black hole. Would that alter the consequences of the thought experiment described in the previous section? By classical physics alone, the answer is negative. Consider the time-reversed evolution of a black hole. Numerous particles must be directed towards one point, time-reversing the black hole's evaporation. Now, {\it even if we allow the information to survive in these particles so as to re-create the in-falling objects inside the black hole, this would not suffice for the emergence of macroscopic objects from the singularity}. This is forbidden by relativity theory. Here too, the time reversal of black hole evaporation fails to produce a white hole. Neither can the time-reversal of Hawking's semi-classical account do that.

Is there some yet-unknown physical law that would give rise, merely by the appropriate choice of positions and momenta of thermal particles, to the exotic solution of Schwarzschild's equation? If such a law exists, it must somehow connect thermodynamics, general relativity and quantum mechanics. This should not surprise us, since Bekenstein's Generalized Second Law (see \cite{Bek}) has indicated such a connection long ago. A much more plausible assumption, however, is Penrose's hypothesis: When quantum gravity is available, it might reveal a fundamental irreversibility at Nature's most fundamental level.

\section {Causality, Time-Asymmetry, and Time's ``Passage"}

In section 2 we noted an observational paradox: no observation or experiment can rule out the possibility that our universe's entropy is decreasing rather than increasing. Now, however, we are in a position to resolve this paradox:

{\it Unless information is somehow retained in the evaporating black hole's particles, and unless there is an unknown principle that enables these particles to produce a white hole, our universe cannot be a time-reversed one. A single black hole would suffice to ruin such a reversal. Since our universe contains (most probably) numerous black holes, its entropy increase in intrinsic, regardless of its initial state}. 

Whence this intrinsic time arrow that emerges when a black hole evaporates? The answer, we propose, lies in the controversial property of time itself mentioned at the beginning of this paper. Time seems to have an intrinsic ``flow." Yet mainstream physics, as noted above, is almost unanimous in its insistence that time's apparent passage is only illusory. Both Penrose and Hawking adhere to this view. 

Surprisingly, however, it is Hawking's information-annihilation argument that gives the issue a new twist, though opposite to his own time symmetry argument. If black holes annihilate information in the way he proposes, then the universe's present ordered state cannot be determined by whatever arrangement of particles in the future. One black hole within our future light-cone suffices to destroy the ability of future arrangements to determine the present. This annihilation of information does not hinder, however, the transformation of the present order into future entropy. Ergo, {\it there is an objective sense in which past events are the causes of future events - and not vice versa}. 

Now, once we recall that the universe's history is {\it replete} with black holes, any time-reversed version of the Second Law would boil down to sheer magic: Why is it that order steadily increases towards the past, despite numerous information-erasing black holes? For any ``atemporal" interpretation of the past as the ``effect" of the future, this steady increase of order is no less that a series of miracles. Recall that Hawking ascribes the difference between past and future to boundary conditions. But if black holes destroy information, then each black hole introduces a new boundary condition into the universe's history. Why, again, does order {\it increase} towards the past after each boundary condition, in perfect chronological gradation? The only way to favor causality over miracles is to grant causality itself an intrinsic direction in time: Entropy increase only towards the future because {\it causation itself} proceeds only from past to future. 

An adherent of the Block Universe might concede that causality is asymmetric and still dismiss time's passage as illusion. While logically such a position is tenable, we find it extremely superfluous. For most physicists, the need to preserve the symmetry of causality is one of the {\it reasons} for preferring the Block Universe. If one admits that future events cannot affect present events, there is no need to believe that these future events ``coexist" with present events. If causality turns out to be asymmetric, it would be much more parsimonious to explain it by simply accepting what our immediate experience keeps telling us: Events affect one another only in the forward time direction because they come into being in that order. 

Whether black holes do indeed destroy information is still an open question. Our proof, however, holds for any theory or model that invokes indeterminism: Given even a single truly random interaction, anywhere in the universe, the observed entropy gradient cannot be explained atemporally. There is an objective sense in which low-entropy events give rise to high entropy events, not {\it vice versa}. Past begets future, not {\it vice versa}.

Common sense greets this conclusion with relief, as it rids us from the absurdity of believing that whatever one chooses to do is determined by some incredible conspiracy of numerous distant particles in one's farthest future.\footnote{ Price \cite{Pri2} claims that the atemporal view does not contradict free will. His argument, in our opinion, is highly unconvincing.} Our intrinsic time-arrow also accords with the yet unexplained CP violation exhibited by neutral kaons, which, by CPT invariance, entails a basic T violation too. ``It is hard to believe," says Penrose (\cite{Pen}, p. 583), ``that Nature is not, so to speak, 'trying to tell us something' through the results of this delicate and beautiful experiment." If quantum gravity, underlying the dynamics of black holes, indeed produces information-annihilating effects, it might reaffirm Penrose's suspicion that time-asymmetry lies at the very foundations of physical law.

\section {Acknowledgments}

It is a pleasure to thank Yakir Aharonov, Jacob Bekenstein, Daniel Rohrlich, and Meir Hemo for illuminating discussions.

\bibliographystyle{unsrt}

\begin{thebibliography}{31}

\bibitem{Haw1} S. W. Hawking, The no boundary condition and the arrow of time, in: Physical Origins of Time-Asymmetry, eds. J. J. Halliwell, J. P\a'erez-Mercader, and W. H. Zurek, 346 (Cambridge University Press, Cambridge, 1994). 
\bibitem{Haw2} S. W. Hawking, Phys. Rev. {\bf D 13} (1976) 191.
\bibitem{Hor} L. P. Horwitz, R. I. Arshansky, \& A. C. Elitzur, Found. Phys., {\bf 18}, (1988) 1159.
\bibitem{Pri} I. Prigogine, From being to becoming: Time and complexity in the physical sciences.(W. H. Freeman \& Co., San Francisco, 1980).
\bibitem{Dav} P. C. W. Davies, About time: Einstein's unfinished revolution (Simon \& Schuster, New York, 1995).
\bibitem{God} K. G\"odel, in: Albert Einstein: Philosopher-scientist, Vol II, ed. P. A. Schillip, 557 (Open Court, La Salle, Ill., 1949).
\bibitem{Oak1} L. N. Oaklander, \& Q. Smith, The new theory of time (Yale University Press, New Haven, 1994)
\bibitem{Oak2} L. N. Oaklander, \& Q. Smith, Time, change and freedom (Routledge, New York, 1995).
\bibitem{Eli} A. C. Elitzur \& S. Dolev, Phys. Lett. {\bf A 251}, (1999) 89.
\bibitem{Dav2} P. C. W. Davies, The physics of time asymmetry (Surrey University Press, London, 1974).
\bibitem{Hal} J. J. Halliwell, J. P\a'erez-Mercader and W. H. Zurek, eds. Physical origins of time-asymmetry (Cambridge University Press, Cambridge, 1994).
\bibitem{Sav} S. F. Savitt, ed. Time's arrow today (Cambridge University Press, Cambridge, 1995).
\bibitem{Zeh} H. D. Zeh, The physical basis of the direction of time. (Springer-Verlag, Berlin, 1989).
\bibitem{Pen} R. Penrose, Singularities and time-asymmetry, in: General relativity: An Einstein Centenary Survey, eds. S. W. Hawking, and W. Israel, 581 (Cambridge University Press, Cambridge, 1979).
\bibitem{Haw3} S. W. Hawking and R. Penrose, The nature of space and time (Princeton University Press, Princeton, 1996).
\bibitem{Haw4} S. W. Hawking, Nature {\bf 248} (1974) 30.
\bibitem{Haw5} S. W. Hawking, Phys. Rev. {\bf D 13} (1976) 191.
\bibitem{Ghi} G. C. Ghirardi, A. Rimini, \& T. Weber, Phys. Rev. {\bf D 34} (1986) 470.
\bibitem{Eli2} A. C. Elitzur, J. Theor. Biol. {\bf 168} (1994) 429.
\bibitem{Dol} S. Dolev, and A. C. Elitzur, Einstein Quart.: J. Biol. Med. {\bf 15} (1998) 24.
\bibitem{Pri2} H. Price, Times arrow and Archimedes' point (Oxford University Press, Oxford, 1996).
\bibitem{Pri3} H. Price, Nature {\bf 340} (1989) 181.
\bibitem{Whe} J. A. Wheeler and R. P. Feynman, Rev. Mod. Phys. {\bf 21} (1949) 425.
\bibitem{Bel} J. S. Bell, Speakable and unspeakable in quantum mechanics (Cambridge University Press, Cambridge, 1987). 
\bibitem{Eli3} A. C. Elitzur, Phys. Lett. {\bf A 167} (1992)  335.
\bibitem{Eli4} A. C. Elitzur, Astrophys. Space Sci., {\bf 244}, (1996) 313. 
\bibitem{Pen2} R. Penrose, Shadows of the mind (Oxford University Press, Oxford, 1994).
\bibitem{Eli5} A. C. Elitzur and L. Vaidman, Found. Phys. {\bf 23} (1993)  987.
\bibitem{Unr} W. Unruh, Time, gravity, and quantum mechanics, in: Time's arrow today, ed. S. F. Savitt, 23 (Cambridge University Press, Cambridge, 1995).
\bibitem{Bek} J. D. Bekenstein, Phys. Rev. {\bf D 7}, 2333 (1973). 
\bibitem{Per} J. Preskill, Do black holes destroy information? In: International symposium on black holes, membranes, wormholes, and superstrings, eds. S. Kalara and D. Nanopoulus 22 (World Scientific, River Edge, NJ, 1993).

\end{thebibliography}

\end{document}